\documentclass[
    ,final            
  ]
  {aipproc}

\layoutstyle{8x11single}

\begin{document}

\title{Novel technique for monitoring the performance of the LAT instrument on board the GLAST satellite}


\author{D.Paneque}{
  address={Stanford Linear Accelerator Center (SLAC)}
}

\author{A. Borgland}{
  address={Stanford Linear Accelerator Center (SLAC)}
}

\author{A. Bovier}{
  address={Stanford Linear Accelerator Center (SLAC)}
}

\author{E. Bloom}{
  address={Stanford Linear Accelerator Center (SLAC)}
}

\author{Y. Edmonds}{
  address={Stanford Linear Accelerator Center (SLAC)}
}

\author{S. Funk}{
  address={Stanford Linear Accelerator Center (SLAC)}
}

\author{G. Godfrey}{
  address={Stanford Linear Accelerator Center (SLAC)}
}

\author{R. Rando}{
  address={INFN/Universita di Padova}
}

\author{L. Wai}{
  address={Stanford Linear Accelerator Center (SLAC)}
}

\author{P. Wang}{
  address={Stanford Linear Accelerator Center (SLAC)}
}

\author{on behalf of the GLAST/LAT collaboration}{
  address={\url{http://www-glast.stanford.edu/cgi-bin/people}}
}

\begin{abstract}
 The Gamma-ray Large Area Space Telescope (GLAST) is an observatory designed to perform gamma-ray astronomy in the energy range 20 MeV to 300 GeV, with supporting measurements for gamma-ray bursts from 10 keV to 25 MeV. GLAST will be launched at the end of 2007, opening a new and important window on a wide variety of high energy astrophysical phenomena . The main instrument of GLAST is the Large Area Telescope (LAT), which provides break-through high-energy measurements using techniques typically used in particle detectors for collider experiments. The LAT consists of 16 identical towers in a four-by-four grid, each one containing a pair conversion tracker and a hodoscopic crystal calorimeter, all covered by a segmented plastic scintillator anti-coincidence shield. The scientific return of the instrument depends very much on how accurately we know its performance, and how well we can monitor it and correct potential problems promptly. 

We report on a novel technique that we are developing to help in the characterization and monitoring of LAT by using the power of classification trees to pinpoint in a short time potential problems in the recorded data. The same technique could also be used to evaluate the effect on the overall LAT performance produced by potential instrumental problems.
\end{abstract}

\maketitle


\section{Methodology: usage of classification trees to compare 2 data sets}

The scientific return of the LAT instrument depends on how 
accurately we monitor its performance, and how promptly we can fix problems.
Such data monitoring is a very complex task, since the LAT contains more 
than 850000 channels in the trackers, 1536 CsI crystals 
and 97 ACD plastic scintillator tiles and ribbons.
The standard way is to monitor the parameter values, correlations and time evolution 
(by means of histograms and charts), to check their consistency with a well known reference. 
This methodology is explained elsewhere \cite{StandardMonitoring}. 

A different (and complementary) approach is to try to find differences between the reference 
data set and the just taken data set; both data sets being represented in a N-dimensional space 
of N selected parameters. Classification trees can provide an efficient way of finding 
potential differences between data sets in an automated fashion. Here we used the Random Forest method 
\cite{RFBreiman}, and a custom interface described in \cite{RRando}.

In this approach, we use the classification error 
to quantify the magnitude of the differences between the two data sets; and we use the 
Z-score value to pinpoint the parameters where the differences lie. 
Both classification error and Z-scores are estimated during the growing of the forest, 
using the so-called Out-Of-Bag (OOB) events, which are a bootstrapped sample of events that were not 
used in the growing of the individual trees of the forest.
The classification error \texttt{OOB Err} is  the percentage of OOB events that were incorrectly predicted by 
the forest. In case of equal data sets (no separation possible): OOB Err $\sim$ 50\%.
If the two event classes can be separated (they are different in some way), then OOB Err $<$ 50\%.
The Z-score is a statistical measure, that relies on the OOB Err, to estimate
the importance of a given variable to distinguish between the two data sets. 
The Z-score quantifies the statistical significance ($\sim$ number of sigmas) 
of the differences between the two event classes in a given parameter. 
Each of the N parameters used to grow the forest has its own Z-score value. 
High Z-score (e.g. $>$ 5) implies large (statistically significant) differences between the 
two event classes in that variable.

\section{Illustration of working principle: quick detection of anomolous data sets}
\vspace{-0.3cm}

In order to test the working principle of this novel technique we chose several 
data sets (event classes) taken during the pre-launch tests during the fall 2006 at the Naval Research Lab. 
These data (Cosmic Rays, mostly muons) were processed with the standard LAT event reconstruction software.
We defined the event \texttt{class A} as 
taken when LAT was supposedly working correctly; this is our reference data set. 
\texttt{Class B} will be the data that needs to be evaluated. The event \texttt{class B1} contains data 
taken when LAT was supposedly working correctly, while \texttt{class B2} is 
data taken when LAT was NOT working correctly. In the B2-type data, the information from half 
layer 0 from tracker tower 10 was not properly read; 
thus there is missing information in some events. Therefore, in this test, we expect to have 
compatibility between A and B1; while we expect differences between A and B2.

Two forests of trees were grown; one using A and B1 type data (A-B1), and the other one using 
A and B2 type data (A-B2). For this test, we used only high level data
(derived quantities, using the reconstruction software, from basic detector outputs)
and we only considered non-empty events which triggered tower 10. 
The random forests were grown using 10000 events, 1000 trees, 80 variables and 4 variables/node. 
The time required to grow each of these forests was less than 1/2 hour in 
dual 1.8GHz Opteron CPU machine. 

\begin{figure}
  \includegraphics[height=.21\textheight]{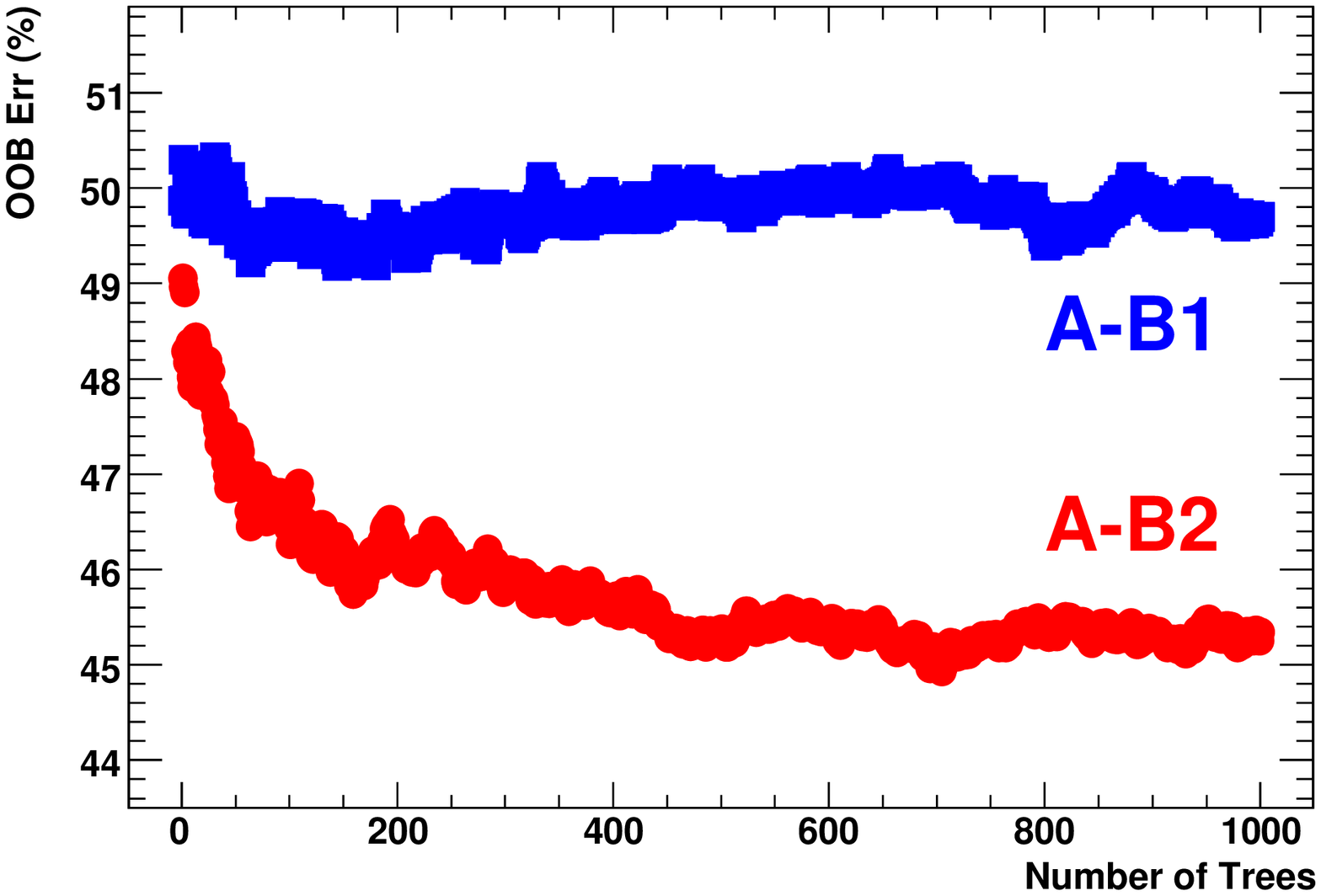}
  \includegraphics[height=.21\textheight]{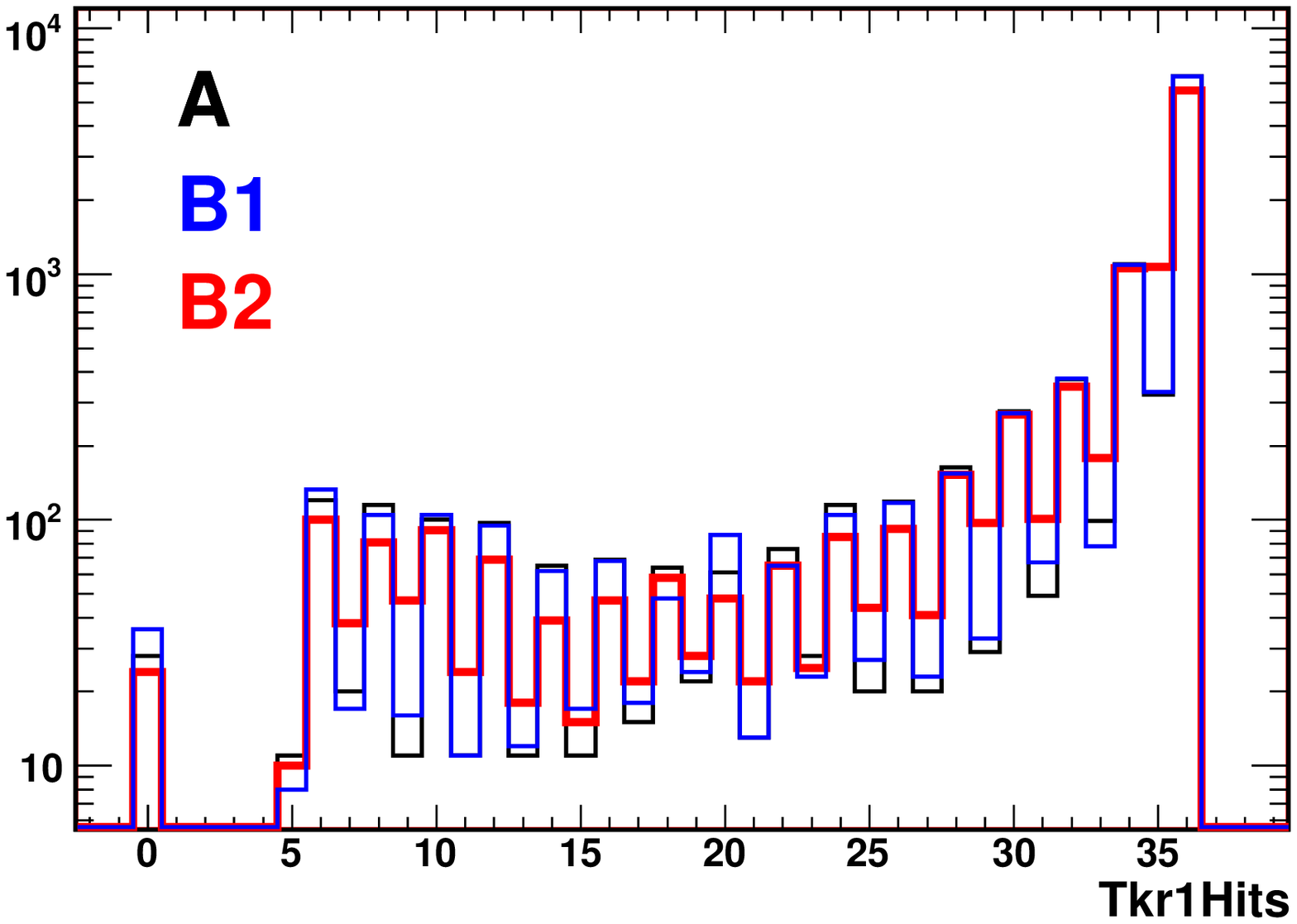}
  \caption{\texttt{Left-hand}; Classification error for A-B1 (blue) and A-B2 (red) event class comparison. 
\texttt{Right-hand} Distribution of Tkr1Hits for the event class A (Black), B1 (blue) and B2 (red, filled histogram).}
\label{Fig1}
\end{figure}

The classification error vs the number of trees is shown in the left-hand plot of Fig. \ref{Fig1}. 
While there is no effective separation between event types A and B1 (Err $\sim$ 50\%),  
the separation between event types A and B2 is clearly possible, 
which implies differences between these event classes. Note also that 
100 trees are enough for a good separation (in this example) which would allow 
us to grow the forest 10 times faster.

The highest Z-score in the A-B2 Random Forest was for the parameter 
that denotes the number of clusters (hit planes) in the main track; Tkr1Hits.
The Z-score for this parameter was 41; which implies large differences in this 
variable. A charge particle passing through all the 19 layers (36 planes) of the LAT 
tracker (all towers) will have Tkr1Hits $\sim$ 36.
The right-hand plot of Fig 1 shows the distribution of Tkr1Hits for the event classes 
A, B1 and B2. Class B2 has a larger fraction of events with odd number of hit planes. 
This is due to the missing information from plane 0 for some of the events.

\vspace{-0.75cm}
\section{Conclusions}
\vspace{-0.25cm}

Random Forest can be a useful tool to monitor the performance of 
LAT during on-orbits operations. A test with pre-launch data suggests 
that the method is fast and efficient. Application of this method to 
low level data would increase the potential of discovering hardware problems, 
at the expense of more computing power.

Note that the application of this method to monitor LAT data during on-orbits
operations is not straight forward. The success depends on: \texttt{a)} the correct selection of the 
reference data set; and \texttt{b)} the selection of the variables (high/low level) and filters
to be used. These selections will be tuned up prior to launch; 
yet this learning will probably continue during the first months of space operation.


 



\vspace{-0.75 cm}
\begin{thebibliography}{9}

\bibitem{StandardMonitoring}
A. Borgland et al., This conference


\bibitem{RFBreiman}
L.Breiman, A. Cutler, 
\url{http://www.stat.berkeley.edu/~breiman/RandomForests/cc_home}

\bibitem{RRando}
R. Rando, 
\url{http://sirad.pd.infn.it/glast/ground_sw/rForest/rforest.html}



\end{thebibliography}
\end{document}